\documentstyle[aps]{revtex} 
\begin{document}
\draft
\title{THE LONDON PENETRATION DEPTH OF STRONGLY COUPLED ISOTROPIC
 SUPERCONDUCTORS:LOW TEMPERATURE BEHAVIOUR }
\author{X.\ Leyronas and R.\ Combescot}
\address{Laboratoire de Physique Statistique, Ecole Normale Sup\'erieure,
24 rue Lhomond, 75231 Paris Cedex 05, France}
\date{Received \today}
\maketitle

\begin{abstract}
We proceed to a systematic exploration of the low temperature
dependence of the London penetration depth of isotropic superconductors
within strong coupling theory in the clean limit. For a sizeable
range of parameters, we find that strong coupling effects can
reasonably simulate a power law dependence, sometimes with an
excellent precision. In such cases it would be quite difficult
to distinguish experimentally between a pure power law and the
strong coupling result. Physically we have been able to ascribe
this temperature dependence to low frequency phonons which produce
a quasi elastic scattering for electrons. The presence of these
low frequency phonons requires rather wide phonon spectra and
their effectiveness in scattering implies fairly strong coupling. 
\end{abstract}

\section {INTRODUCTION} 
The temperature dependence of the London penetration depth 
$\lambda _{L}(T) $   
plays an important role in the ongoing debate about the mechanism  
of high $T_{c } $ superconductivity. It is indeed an essential  
piece of information about this mechanism to know the symmetry  
of the order parameter \cite{sd}. While an s-wave type order parameter  
( that is having a fixed sign over the Fermi surface ) finds  
a natural explanation in terms of a purely attractive interaction,  
an order parameter which changes sign over the Fermi surface,  
such as the d-wave one produced by the spin fluctuation mechanism,  
points toward some repulsive component in the pairing interaction.  
A change of sign for the order parameter implies the existence  
of nodes of the gap and hence of low energy elementary excitations  
in the superconducting phase. These should manifest themselves  
experimentally in numerous low temperature properties. However  
the interpretation of  many experiments is not so easy. In this  
respect the penetration depth is a very interesting quantity  
to measure since it is a thermodynamic quantity and its measurement  
is less likely to be perturbed by extrinsic defects than a dynamical  
quantity. Moreover it is a bulk quantity since one explores  
the sample over typically 1000 \.A, in contrast to photoemission  
or tunnelling where only a few atomic layers at the surface  
are involved, which raises the fear that the surface might be  
perturbed or perturbing in some way.   \par 
  \bigskip 
The existence of low energy excitations should 
produce for $\lambda _{L }(T) $   
a low temperature  dependence stronger than the standard BCS  
behaviour, which is essentially flat. Therefore many experimentalists  
have looked at this low T behaviour. In particular they have  
quite often tried to fit this behaviour with a power law $T^{n } $  
since this would allow to find if the nodes of the gap, if any,  
are located on points or lines on the Fermi surface. Indeed  
many experiments on YBCO and BiSSCO \cite{litexp} have found near  
$T^{2} $  behaviours which have been interpreted as proving the  
existence of point nodes. More recently YBCO cristals \cite{hardy}  
and films \cite{lanb} have shown a T behaviour, pointing toward lines  
of nodes at the Fermi surface. There is however an implicit  
assumption in this kind of conclusion. This is that no s-wave  
superconductor can give a low temperature dependence compatible  
with experiment. This assumption has been mainly challenged  
by pointing out that strongly anisotropic s-wave superconductors  
will clearly give a low T behaviour stronger than standard BCS.  
A related point is that normal layers will also produce low  
energy excitations and a strong low T dependence \cite{klemm}.   \par 
  \bigskip 
However even for a completely isotropic superconductor there  
is also another direction to look for possible physical effects  
which would spoil the standard interpretation. Indeed, when  
comparison is made with experiment, the standard reference theory  
for s-wave superconductivity is the weak coupling BCS theory.  
On the other hand there are good indications that high $T_{c } $  
 superconductors might be in the strong coupling regime. Indeed,  
in addition to the high value of $T_{c } $  itself, resistivity as  
well as infrared experiments give for the inverse quasiparticle  
lifetime at $T_{c } $    a value which is of order of $T_{c } $  itself,  
in clear contradiction with the weak coupling assumption which  
requires it to be much smaller than $T_{c }. $ It is therefore  
worthwhile to wonder about strong coupling effects on the penetration  
depth at low temperature. Although much work \cite{carb}has been  
devoted to the strong coupling theory of $\lambda _{L }(T) $ and qualitative   
suggestions have been made on strong coupling effects \cite{nemelia},  
there has been to our knowledge no quantitative and systematic  
study of the low temperature behaviour. This is naturally easy  
to understand since, before the discovery of high $T_{c } $ 
superconductivity and the recent debate on the symmetry, there was no specific  
reason to look in details at this specific point. It is our  
purpose in this paper to consider this theoretical question  
in detail, both as a basic exploration in strong coupling theory  
and as a reference for the interpretation of experiments in  
high $T_{c } $  compounds. We will consider only isotropic superconductors.  
Our purpose is indeed to show that even in this case, the low  
temperature behaviour can be qualitatively different from the  
BCS result. Naturally strong anisotropy will produce important  
additional changes at low temperature. We will come back on  
this point at the end of the paper. In the next section we give  
our procedure for this systematic exploration. In section III  
we give our results and discuss their physical origin. Finally  
we discuss in the last section the consequences of our findings.  \par

\section {POSITION OF THE PROBLEM}
The fact that the penetration depth is a thermodynamic quantity  
is an important advantage for the study of strong coupling effects.  
Indeed the input of any calculation in the strong coupling Eliashberg  
theory is the so-called Eliashberg function $\alpha ^{2}F(\Omega ) $ which  
gives the strength of the electron phonon interaction for a  
given phonon frequency $\Omega  $ ( naturally we might consider as well  
an attractive pairing interaction caused by the exchange of  
other kinds of bosons, but in the following we will consider  
phonons for simplicity ). In principle the knowledge of 
$\alpha ^{2}F(\Omega ) $   
is equivalent to fixing an infinite number of parameters. In  
practice this forces to consider a set of models for $\alpha ^{2}F(\Omega ) $  
depending on a restricted number of parameters. These models  
are in general either inferred from experimental data or chosen  
for various theoretical reasons. Nevertheless this procedure  
is not satisfactory since the chosen models contain necessarily  
some arbitrary ingredients : either these are irrelevant and  
it would be better to know it, or they are important in which  
case they are not under control.   \par 
  \bigskip 
However it has been pointed out recently that it is possible  
to overcome this difficulty for thermodynamic quantities \cite{gaptc}.  
Indeed the very structured Eliashberg function $\alpha ^{2}F(\Omega ) $  
enters actually the imaginary axis Eliashberg equations only 
through the very smooth spectral function $\lambda (\omega ) 
$ = $\int  $ $d\Omega  $ $2\alpha ^{2}F(\Omega ) $ 
$\Omega /(\Omega ^{2}+\omega ^{2}) $  
which gives physically the frequency dependence of the effective  
pairing interaction. It has been shown \cite{gaptc}that it is possible  
in a systematic way to approximate very well ( within a few  
percent ) all the possible $\lambda (\omega ) $ by a set of functions, 
depending   on 5 parameters, which provide in this way a ''representation''  
of all the possible spectra $\alpha ^{2}F(\Omega ) $.
 For the critical temperature   
and the gap, it was indeed found that the difference between  
the quantity calculated from the original spectrum and from  
its representation differed at most by a few percent. Although  
there are clearly many different ways to choose the functions  
which give the representation, we have found it convenient to  
use all the functions generated by a spectrum made of two Einstein  
peaks \cite{gaptc}. In this case the five parameters are, in addition  
to the Coulomb pseudopotential $\mu *, $ the frequencies $\Omega _{1} $  
and $\Omega _{2} $  of the two peaks and their  weights $\lambda _{1} $  and  
$\lambda _{2} $  , where $\lambda _{1} $  + $\lambda _{2} $  = $\lambda  $ 
with $\lambda  $ being the  
total coupling strength. We set  r = $\Omega _{2} $    / $\Omega _{1} $  and  
$\rho  $ = $\lambda _{2} $  / $\lambda _{1}. $ These parameters are 
obtained from   
the original spectrum \cite{gaptc} by   
r = ( 1 + $b\rho ^{-1/2} $ )/(  1 $- $  $b\rho ^{1/2}) $ , 
where 1 + $b^{2} $ = $< $ $\Omega ^{2} $ $>/< $ $\Omega  $ 
 $>^{2} $  
. And $\rho  $ is the solution of  $r^{-\rho /(1+\rho )} $  ( 1 + $r\rho  $  
)/( 1 +  $\rho ) $ = $<\Omega >/ $ $\Omega _{log} $ , 
which is easily solved numerically.  
Here $< $ $\Omega ^{\beta } $ $> $ = $\int  $ $d\Omega  $ 
$\Omega ^{\beta } $  $2\alpha ^{2}F(\Omega )/\lambda \Omega  $ 
and $\Omega _{log} $  
 = lim $_{\beta  \rightarrow  0} $ 
$<\Omega ^{\beta } $ $>^{1/\beta } $ .
We stress that the  
exploration of this representation will allow us to cover all  
the possible cases arising in strong coupling theory.   \par 
  \bigskip 
In the present case the situation simplifies somewhat since  
we will consider the variation of the reduced penetration depth  
$\lambda _{L }(T) $ / $\lambda _{L }(0) $  as a function of the reduced 
temperature  
T / $T_{c }. $ Therefore the absolute values of the frequencies  
will naturally be irrelevant and, in addition to $\mu *, $ we are  
left with only 3 parameters to vary, namely the total coupling  
strength $\lambda , $ the ratio  r  of the peak frequencies which is  
a measure of the width of the spectrum, and the relative weight  
p = $\lambda _{2} $  / $\lambda  $  of the high frequency peak compared to   
the total coupling strength. Actually we will consider only  
the case $\mu * $ = 0 in this paper although we have also explored  
the case $\mu * $ = 0.1 . The reason for this will be clear in the  
following. Indeed the Coulomb pseudopotential is a high frequency  
contribution to the pairing interaction while we will see that  
all the interesting physics comes from low frequencies, as we  
might expect at low temperature. Therefore $\mu * $ provides essentially  
an uninteresting renormalization of the coupling strength.  \par 
  \bigskip 
Our endeavour meets from the start with a basic problem, which  
is both conceptual and practical. Whereas it is straightforward 
to give the variation  of a physical quantity as a function 
of various parameters,  
how can we study the low temperature dependence of $\lambda _{L }(T) $  
as a function of these parameters ? Indeed we do not know of  
any general functional dependence of $\lambda _{L }(T) $ in this regime,  
which would then depend only on a few parameters. In fact this  
general problem is analogous to the one raised above for 
$\alpha ^{2}F(\Omega ) $   
and the solution should be the same : we ought to expand $\lambda _{L }(T) $   
on an appropriate functional basis and find the coefficients  
of this expansion. Actually the present experimental situation  
and the way most results are analyzed suggest a natural way  
to do something of this kind. We will try to find an approximate  
power law dependence for $\lambda _{L }(T) $ at low temperature, which  
provides us with an exponent and a prefactor to characterize  
this behaviour. Naturally we will find in general that $\lambda _{L }(T) $   
does not obey a power law. Therefore we will proceed specifically  
in the following way. We work with the superfluid density 
$\rho _{s }(T) $   
= $\lambda ^{-2}_{L }(T) $ / $\lambda ^{-2}_{L }(0) $ and define apparent   
exponents between temperature $T_{i } $  and $T_{f } $  as  n = ln  
[( 1 - $\rho _{s }(T_{f }) $ ) / 
( 1 - $\rho _{s }(T_{i }) $ ) ] 
/ ln $(T_{f } $  
/ $T_{i } $ ) . Both for experimental and theoretical reasons  
we do not take our sampling temperature too low. Indeed for  
the experiments we are interested in \cite{remarque}, the low temperature  
behaviour is pretty flat with some scatter and the resulting  
imprecision makes results obtained in this region rather meaningless.  
On the theoretical side we could naturally go in this region,  
but the variations of $\rho _{s}(T) $ would be minute. Moreover we  
know theoretically that a power law will be a very bad description  
of $\lambda _{L }(T) $ in this regime. Therefore we have chosen our sampling   
temperatures to cover the ''middle-low'' temperature region 0.2  
$< $ T / $T_{c } $  $< $ 0.5 where an experimentalist is most likely  
to look for a power law. Precisely we have taken  $T_{1} $  /  
$T_{c } $  = 0.2,  $T_{2} $  / $T_{c } $  = 0.35 and  
$T_{3} $  / $T_{c } $   
 = 0.5, and we have defined three exponents : $n_{1} $  between  
0.35 and 0.5 , $n_{2} $  between 0.2 and 0.5, and $n_{3} $  between  
0.2 and 0.35 (see Fig.1). Naturally $n_{2} $  = 0.389 $n_{1} $   
+ 0.611 $n_{3} $ , but we find it convenient to use this redondant  
presentation because  $n_{3} $ - $n_{1} $ appears as a kind of theoretical  
''error bar'' for the middle exponent $n_{2}. $ A large value will  
tell us that a power law is a bad representation of the temperature  
dependence of $\lambda _{L }(T) $ in this region, while a small error  
bar will mean that a simple power law provides a good description.  
Most of the time we have found $n_{1} $  $< $ $n_{2} $ $< $ $n_{3} $  
corresponding   to a weaker T dependence at lower temperature, but it 
is important  
that the reverse order occurred also. Naturally our description  
could be improved, for example by replacing the prefactor by  
a polynomial, but we have not explored this possibility.  \par

\section{RESULTS ON THE PENETRATION DEPTH} 
The London penetration depth is given by \cite{Nam}:  

\begin{eqnarray}
{\lambda }_{L}^{-2}(T)={\lambda }_{L}^{-2}2\pi
T\sum\nolimits\limits_{n=0}^{\infty } {{\Delta }_{n}^{2} \over
{\omega}_{n}^{2}+{\Delta }_{n}^{2}}{1 \over {Z}_{n}({\omega }_{n}^{2}+
{\Delta}_{n}^{2}{)}^{1/2}+{1 \over 2{\tau }_{i}}}
\label{eq1}
\end{eqnarray}
Here $\lambda _{L } $  is the weak coupling penetration depth at T =  
0  in the clean limit, given by $\lambda _{L }^{-2} $ $\equiv  $ (2/D) 
$N_{0} $  
$e^{2} $ $v_{F }^{2} $   where D is the dimensionality of the 
superconductor,$N_{0} $ the density of states per spin at the Fermi level,
$\tau _{i } $  is the impurity scattering time, 
and  $\Delta _{n } $  and   $Z_{n } $    are respectively the gap 
function and the phonon  
renormalization function at the Matsubara frequencies $\omega _{n } $  
= $(2n+1)\pi T. $ These last ones are given by the solution of the  
Eliashberg equations:  

\begin{eqnarray}
{\omega }_{n}({Z}_{n}-1)=\pi T\matrix{\cr
\sum\nolimits\cr
m\cr}{\lambda }_{n-m}{{\omega }_{m} \over ({\omega }_{m}^{2}+
{\Delta }_{m}^{2}{)}^{1/2}} \\
{\Delta }_{n} {Z}_{n}=\pi T\matrix{\cr
\sum\nolimits\cr
m\cr}{\lambda }_{n-m}{{\Delta }_{m} \over ({\omega }_{m}^{2}+
{\Delta}_{m}^{2}{)}^{1/2}} \nonumber
\label{eq2}
\end{eqnarray}
where:  

\begin{eqnarray}
{\lambda (\omega )}^{}=\lambda <{{\Omega }^{2} \over {\Omega }^{2}+
{\omega}^{2}}>
\label{eq3}
\end{eqnarray}
with  $\lambda  $ $\equiv  $ $\lambda (0). $ As above the 
brackets   $<...> $  are for the  
average  $\int  $ $d\Omega  $ $g(\Omega ) $ ...  over phonons 
frequency with respect   to the normalized Eliashberg function 
$g(\Omega ) $ = $2\alpha ^{2}F(\Omega )/\lambda \Omega  $   .   \par 
  \bigskip 
Although this imaginary axis expression Eq.(1) is by far the  
most convenient for numerical calculations, it is useful for  
the physical interpretation of the results to write also the  
corresponding expression with integration on the real frequency  
axis. Indeed this allows to express the penetration depth, which  
gives the superfluid response to the electromagnetic field,  
in terms of virtual pair-breaking excitations. This expression  
reads: 

\begin{eqnarray}
{\lambda }_{L}^{-2}(T)={\lambda }_{L}^{-2}Im\int_{0}^{\infty }d\omega
tanh({\omega  \over 2T}) {{\Delta }^{2}(\omega )
 \over ({\Delta}^{2}(\omega )-{\omega }^{2}{)}^{3/2}}{1 \over {Z(\omega )}
+{1\over 2{\tau }_{i}({\Delta }^{2}(\omega )-{\omega }^{2}{)}^{1/2}}}
\label{eq4}
\end{eqnarray}
It reduces to Eq.(1) by deforming the integration contour toward  
the imaginary frequency axis. This result is easy to understand  
physically. If one omits the last term in the integral, this  
expression is the standard BCS result except for the frequency  
dependence of $\Delta (\omega ). $ The last term corresponds to the mass  
renormalization of the excitations. Indeed in addition to the  
constant factor $1/m^{*} $  due to the band structure effective  
mass coming in the prefactor  $\lambda _{L }^{-2} $ , one has to  
include the frequency dependent mass renormalization function  
$Z(\omega ). $ Naturally the imaginary part of  $Z(\omega ) $ will describe  
the effect of the finite lifetime of the excitations. The additional  
contribution  $(1/2\tau _{i }) $ $(\Delta ^{2}(\omega ) $  - 
  $\omega ^{2})^{-1/2} $   in the last term is just the corresponding 
effect of impurities  
on the excitation lifetime. In this paper we will actually be  
concerned only with the clean limit. Indeed we will see that  
the interesting low temperature behaviour of $\lambda (T) $ is controlled  
by lifetime effects due to inelastic scattering. When the superconductor  
gets very dirty, the elastic lifetime becomes much shorter than  
the inelastic one and the low temperature behaviour is similar  
to the standard weak coupling BCS result and therefore uninteresting  
for our purpose. We note also that, if we have in mind a comparison  
with experiment, the dirty limit is not relevant for high $T_{c } $  
compounds which are more or less on the clean side because $T_{c } $  
is so high. Therefore the clean limit is physically the interesting  
one. However Eq.(1) and (4) will be useful for comparison.   \par 
  \bigskip 
Let us start with the simplest case, namely the Einstein spectrum,  
in order to have a reference for comparison. We display in Fig.2  
the results for an Einstein spectrum, which corresponds to take  
our parameter  p = 0  or  1 , the value of  r  being then irrelevant.  
We do not give the values of our exponents $n_{i } $ for these  
Einstein spectra since they range typically from 5 to 10 and  
a power law is clearly a very poor way to describe the low temperature  
behaviour for them as it is clear from Fig.2. The $\lambda  $ = 0 curve  
is naturally the BCS result. At the beginning, when the coupling  
strength starts to increase, $\lambda _{L }(T) $ becomes flat at even  
higher temperature than in the BCS result. This is just the  
effect of the increase with $\lambda  $ of the ratio 2 $\Delta  $ / $T_{c } $ 
 .  
Indeed, as it is seen from Eq.(4), the flattening of $\lambda _{L }(T) $  
starts when all excitations have essentially disappeared. Since  
the gap $\Delta  $ gives the minimum energy for creating an excitation,  
the temperature at which $\lambda _{L }(T) $ becomes flat scales with  
$\Delta . $ However when we have, say,  $\lambda  $ $> $ 4, 
this evolution is reversed   
and the low temperature dependence gets stronger. When we reach  
$\lambda  $ $\approx 10 $ , we are already back to 
the $\lambda =1 $ behaviour. Since   
we have specifically in mind here the low temperature behaviour,  
it is of interest to fully understand physically the origin  
of this somewhat surprising evolution.   \par 
  \bigskip 
Since this behaviour appears on the strong coupling side, we  
will understand it by considering the strong coupling limit,  
where the coupling constant $\lambda  $ goes to infinity, while we let  
at the same time the phonons frequencies go to zero in order  
to keep for example  $\lambda  $ $< $ $\Omega ^{2} $  $> $  = 1 
in order to keep the  
critical temperature fixed ( here we do not restrict ourselves  
to an Einstein spectrum and consider rather a general spectrum  
). This limit has already been investigated in various papers  
\cite{alldynes,bergrai,marcar,russ,comb,carb}. 
However for our purpose it is more  
interesting not to let $\Omega  $ go strictly to zero, but to take it  
very small in order to avoid singularities 
at low temperature. Because of the factor $\lambda _{n -m } $ , the  
dominant contribution in the sums in Eq.(2) comes from the terms  
with  $|\omega _{m } $ - $\omega _{n }| $  at most of order 
a few times $\Omega . $   
When $\Omega  $ becomes very small,  $\omega _{m } $  $(\omega _{m }^{2} $ + 
$\Delta _{m }^{2} $  
$)^{-1/2} $  is almost constant over this range since $\Omega  $ will  
be very small compared to $\Delta _{m }. $ Therefore the sum can be  
performed explicitely which leads, to leading order, to:    

\begin{eqnarray}
{Z}_{n}={\pi  \over 2}{\lambda }^{}<\Omega (2N(\Omega )+1)>
{1 \over ({\omega}_{n}^{2}+{\Delta }_{n}^{2}{)}^{1/2}}
\label{eq5}
\end{eqnarray}
where $N(\Omega ) $ is the Bose - Einstein distribution. The analytic  
continuation of this simple result toward the real axis :  

\begin{eqnarray}
{Z}^{}(\omega )=i{\pi  \over 2}{\lambda }^{}<\Omega (2N(\Omega )+1)>
{1 \over({\omega }^{2}-{\Delta }^{2}(\omega ){)}^{1/2}}
\label{eq6}
\end{eqnarray}
can also be obtained directly from the real axis Eliashberg  
equations. The physical meaning of Eq.(5) and (6) is quite clear.  
The very low frequency phonons behave like quasi-impurities  
\cite{bergrai}. When they are absorbed or emitted, they produce the  
same result as impurities with a scattering time $\tau _{ph} $   
given by  $1/\tau _{ph} $ = $\pi  $ $\lambda  $ $< $ $\Omega  $  
$( $  $2 $ $N(\Omega ) + 1 )>, $  as it can be seen by comparing 
Eq.(4) and (6) ( one  
would get the same result for  $1/\tau _{ph} $  in the normal state  
). In other words in this limit $Z(\omega ) $ is physically dominated  
by lifetime effects. When this result is carried into Eq.(1)  
one finds naturally the dirty limit with  $\tau _{ph} $    
as scattering   time:   

\begin{eqnarray}
{\lambda }_{L}^{-2}(T)={\lambda }_{L}^{-2}{2 \over \pi \lambda <\Omega
coth(\Omega /2T)>}2\pi T\sum\nolimits\limits_{n=0}^{\infty }
{{\Delta }_{n}^{2}\over {\omega }_{n}^{2}+{\Delta }_{n}^{2}}
\label{eq7}
\end{eqnarray}
The last factor, which contains the sum over Matsubara frequencies,  
gives a very regular function of temperature. This function  
is given in \cite{carb}. It goes to a constant \cite{disc} when 
T $\rightarrow  $ 0.  
Therefore the low temperature behaviour of $\lambda _{L }^{-2} $  
(T) is given by $\tau _{ph } $ . As long as $T $  $>> $  $\Omega  $  , 
 $\lambda _{L }^{-2} $  (T) diverges as T $^{-1} $ , because
the number of thermally  excited phonons is proportional to T : 
when the temperature  
is lowered the number of scattering processes decreases which  
produces the growth of $\lambda _{L }^{-2} $  (T). Naturally this  
divergence saturates for $T $  $<< $  $\Omega  $ , since at zero temperature  
only the possibility of spontaneous phonon emission is left  
which gives $1/\tau _{ph } $ = $\pi  $ $\lambda  $ $< $ $\Omega  $    $>. $ 
We come therefore  
to the conclusion that the increased low temperature dependence  
found when the coupling gets very strong is due to the low frequency  
phonons, which act as scatterers and whose number depends naturally  
on temperature.  \par 
  \bigskip 
We turn now to our investigation of the apparent exponent for  
the low T behaviour of 1 - $\lambda ^{-2}_{L }(T) $
 / $\lambda ^{-2}_{L }(0). $   
Since we have already explored the strong coupling limit we  
can restrict ourselves to ''reasonable'' values for the coupling  
constant $\lambda . $ Specifically we consider $\lambda  $ going from 2 to 8.  
Indeed for $\lambda  $ = 1 we have found results rather similar to the  
BCS result with fairly large exponents ( of order 4 or more  
) which have little interest. Similarly we do not show the results  
for a spectral width parameter  r = 2 , since they do not depart  
very much from what is found for Einstein spectra. We consider  
first the results for  r = 4 , which are displayed in Fig.3  
. Since p = 0 or 1 correspond to Einstein spectra, we give only  
the results for 0.1 $\leq  $ p $\leq  $ 0.9 . While for 0.1 $\leq  
$ p $\leq  $ 0.5, it  
is plain that a power law does not agree with the theoretical  
results, remarkably it becomes a reasonable representation for  
0.6 $\leq  $ p $\leq  $ 0.9 , mainly for high values of $\lambda . $ 
Indeed if we accept  
an uncertainty of $\pm  $ 10\% on the exponent ( it is not easy to  
obtain a better result experimentally ) we find a power law  
for 0.7 $\leq  $ p $\leq  $ 0.9 provided $\lambda \geq 3. $  
In particular for p = 0.8  
we find that for $\lambda  $ = 4 ( $n_{i} $ = 3.55 ), 6 ( $n_{i} $ = 2.95   
) and 8 ( $n_{i} $ = 2.72 ) the temperature dependence is remarkably  
well described by a power law. This is somewhat surprising since  
we know theoretically that this dependence is not a power law,  
and there is no obvious reasons why it comes so close to be  
one. We might indeed have anticipated that all our results would  
behave the way they do for, say, p $\leq  $ 0.5 . Naturally the fact  
that $n_{1} $  = $n_{3} $  does not imply that we have exactly a  
straight line on our log-log plot, and we might worry that it  
has actually a sizeable oscillation. One can check directly  
that it is not so : the difference with a straight line is quite  
small and could certainly not be seen experimentally. We give  
right below a specific example of this.  \par 
  \bigskip 
When we look for larger values of the spectral width, we find  
that the range of parameters where a power law gives a good  
description gets larger and that the exponents decrease. This  
is seen on Fig.4 where we present our results for r = 6. For  
0.6 $\leq  $ p $\leq  $ 0.8 we obtain good power laws within $\pm$ 10\%  
down to   $\lambda  $ = 3. For p = 0.75 we find $n_{i} $ = 2.6  
for $\lambda  $ = 4 . This  
exponent is already rather near experimental results. We show  
on Fig.5 the result for 1 - $\lambda ^{-2}_{L }(T) $ / 
$\lambda ^{-2}_{L }(0) $   
compared with an exact power law. It is clear that it would  
not be possible experimentally to make the difference between  
the power law and the strong coupling result for temperature  
below 0.5 $T_{c } $  . For p=0.7 and $\lambda  $ = 8 one finds a $T^{1.98} $   
law. The results for r = 8 displayed on Fig.6 show the same  
trends. The domain where we have a power law within $\pm 10\% $ extends  
now from p = 0.5 to p = 0.8 and starts almost from $\lambda  $ = 2. For  
p = 0.7 and $\lambda  $ = 4 we find an exponent $n_{i} $  = 2.15 . Finally   
one sees on Fig.7 , where we show the results for  r = 16 ,  
that exponents markedly below 2 and actually not so far from  
1 can be obtained. For example for p = 0.6 and $\lambda  $ = 4 we have  
$n_{i} $  = 1.51 . The domain with a power law within $\pm 10\% $ goes  
from p = 0.4 to p = 0.7, starting from $\lambda  $ = 2. We complete our  
results by displaying in Fig.8 the prefactor of the approximate  
power laws that we have obtained for r = 4, 8 and 16 (this prefactor  
is calculated between T / $T_{c } $ = 0.2 and 0.5 using our exponent  
$n_{2} $  ). We give only the results for the values of  p  where  
a power law is a good approximation. The prefactor that we find  
is always of order unity. This means that the power law that  
we find at low temperature is not a small effect since its extrapolation  
for   T = $T_{c } $      gives for 1 - $\lambda ^{-2}_{L }(T) $ / 
$\lambda ^{-2}_{L }(0) $   
a result of order unity.   \par 
  \bigskip 
Naturally the parameters where we find very good power laws  
correspond to the region where one switches from  $n_{1} $  $< $  
$n_{2} $ $< $ $n_{3} $   to  $n_{3} $  $< $ $n_{2} $ $< $ $n_{1} $  
for the order   
of our exponents. We have found above that this crossing occurs  
in the region p $\approx  $ $0.7\pm  $ 0.1 . However since p = 0 or 1 
corresponds   
to an Einstein spectrum, the exponents must be in the same order  
for p = 0 and 1 which implies that there must be another crossing  
region. This region is actually very near p = 1 (for r = 4 and  
$\lambda  $ = 3 it occurs for p = 0.98, and for higher values of r  it  
goes closer to p = 1 ). This means that the spectrum has a very  
small low frequency component, which makes this kind of spectrum  
somewhat pathological and unlikely to be relevant experimentally.  
The exponents corresponding to this crossing are also fairly  
high which makes them quite uninteresting. Therefore we do not  
discuss this region further. In contrast the region which we  
have considered above corresponds to quite standard spectra  
and the exponents are similar to what is found experimentally.  
  \par 
  \bigskip 
It is worthwhile to try to understand the physical origin of  
this appearance of quasi-power laws with small exponents. Since  
this feature develops when the spectral width gets large, it  
is useful to consider the large spectral width limit, which  
corresponds to let go to infinity our parameter  r  in the two  
peaks representation. This is equivalent to let the frequency  
$\Omega _{1} $  of the lower peak go to zero, while we keep at the  
same time the higher frequency peak $\Omega _{2} $  fixed ( we can  
take for example $\Omega _{2} $  as unity). Physically, in the same  
way as we have seen for the strong coupling limit, the low frequency  
phonons will behave as impurities. They disappear from the equation  
for the gap function. Therefore $\Delta _{n } $  as well as $T_{c } $   
are given by the Einstein spectrum result with frequency $\Omega _{2} $  
and coupling strength $\lambda _{2}.{\bf} $  On the other hand the low  
frequency phonons still contribute to $Z_{n } $   by the m = n  
term. This produces the same effect as introducing additional  
scattering with a lifetime  $1/2 \tau _{ph} $ = $\pi  $ $\lambda _{1} T$.
This leads for the penetration depth to:  
 
\begin{eqnarray}
{\lambda }_{L}^{-2}(T)={\lambda }_{L}^{-2}2\pi
T\sum\nolimits\limits_{n=0}^{\infty } {({\Delta }_{n}^{E}{)}^{2}
\over {\omega}_{n}^{2}+({\Delta }_{n}^{E}{)}^{2}}
{1 \over {Z}_{n}^{E}({\omega}_{n}^{2}+
({\Delta }_{n}^{E}{)}^{2}{)}^{1/2}+\pi{\lambda }_{1}T}
\label{eq8}
\end{eqnarray}
where, as explained above, $\Delta _{n }^{E} $  and  $Z_{n }^{E} $  
are the solution of Eliashberg equations for an Einstein spectrum  
with frequency $\Omega _{2} $ and coupling strength $\lambda _{2} $ . As one  
might have expected, this result is just the general formula  
Eq.(1) with the impurity lifetime replaced by the contribution  
from low frequency phonons. It is clear from Eq.(8) that, because  
of the term $\pi \lambda _{1}T, $ the low frequency phonons 
depress $\lambda _{L }^{-2} $  (T) below what would be obtained 
from an Einstein spectrum  
alone. Hence they give for $\lambda _{L } $  $   (T) $ at low temperature   
a dependence which is stronger than the standard BCS one. Therefore  
the low frequency phonons are responsible for development of  
the low temperature dependence when the width of the spectrum  
increases. However there is no obvious reason why this dependence  
can get close to a power law. Nevertheless we note that Eq.(8)  
gives a linear T dependence for $\lambda _{L } $  $   (T) $ 
at low temperature   
(naturally our limiting case corresponds to a situation where  
$\Omega _{1} $  $<< $ T). We show on Fig.9 the result of the calculation  
when the coupling strength of the high frequency peak is equal  
to $\lambda _{2} $  = 3. The coupling strength $\lambda _{1} $  of the 
low frequency   
phonons takes the values m $\lambda _{2} $ /(10 -m) and the integer  
m goes from 0 to 9. The values of $\lambda _{1} $  for m $\geq  $ 6 are quite  
large and are just given to show the trend. It is interesting  
to note that the heavy line on this figure, which corresponds  
to $\lambda _{1} $  = 1.28 , is quite similar to the experimental results  
of Ref.\cite{hardy}.  \par 

\section{DISCUSSION}
In this paper we have explored the effect of strong coupling  
on the low temperature behaviour of the penetration depth in  
the simplest model, namely for an isotropic attractive interaction  
leading to s-wave pairing. Quite surprisingly and unexpectedly  
we have found that strong coupling effects can mimick a power  
law dependence, for a sizeable range of parameters, sometimes  
with an excellent precision. Physically we have been able to  
ascribe this strong T dependence to low frequency phonons which  
produce a quasi elastic scattering for electrons, but we have  
not found any deeper theoretical reasons for these quasi-power  
laws. The presence of these low frequency phonons below $T_{c } $  
 requires rather wide phonon spectra and their effectiveness  
in scattering implies fairly strong coupling.   \par 
  \bigskip 
Let us try to apply our results to the case of high $T_{c } $ 
superconductors.   
Although we have found that one can obtain in principle a linear  
low temperature dependence from strong coupling effects, it  
seems completely unlikely that such an interpretation applies  
to the linear dependence found recently in YBCO \cite{hardy,lanb}  
because our parameters are too extreme in terms of phonon frequency  
and coupling strength. On the other hand our results are of  
interest for other high $T_{c } $ compounds : if we take the experimental  
results showing a dependence not so far from a $T^{2} $  law,  
as it is quite often found \cite{litexp} in BiSSCO, our results are  
rather close to provide an alternative explanation although  
our coupling strength is a bit too high and the spectrum we  
require somewhat too wide to agree with the experimental information  
presently available.  \par 
  \bigskip 
On the other hand we have only investigated here the isotropic  
case. It is clear that, if we consider some anisotropy, the  
effects that we have found will be increased. Indeed, as we  
see from Eq.(8), the inverse lifetime due to low frequency phonons  
is in direct competition with the size of the gap. Anisotropy  
will make the gap smaller at some places on the Fermi surface,  
thereby increasing the effect of low frequency phonons and leading  
to the possibility of having a power law at low T for reasonable  
values of the parameters. In this way strong coupling effects  
might be quite relevant to the understanding of experimental  
results. In this respect it is worth pointing out that most  
high $T_{c } $  microscopic theories claiming to perform a reliable  
critical temperature calculation are strong coupling theories  
with coupling constants at least of order unity. This is in  
particular the case for spin fluctuations microscopic theories  
such as the MMP \cite{pines} or the RULN \cite{ruln}calculations. How  
strong should be the coupling in these compounds is still naturally  
a matter of debate. However strong coupling effects are generally  
completely overlooked in the interpretation of experiments on  
the low temperature behaviour of $\lambda _{L }(T). $ Our results show  
that this might be dangerous. Another way to put it is to say  
that, when something happens at low temperature, one should  
not only look at the fermionic degrees of freedom for an explanation,  
but one should also consider the possibility that bosonic degrees  
of freedom might be at least partly responsible for the effect.  \par 

\section{ACKNOWLEDGEMENTS} 
We are very grateful to N. Bontemps and P. Monod for many discussions  
on the penetration depth question.   \par

\begin{figure}
\caption{Example of the calculation of our apparent power law  
exponents $n_{1}, $ $n_{2} $ and $n_{3} $ from the log-log plot of  
 1 $- $ $\lambda _{L }^{-2} $  (T) / $\lambda _{L }^{-2} $ 
 (0) ( this figure  
corresponds precisely to the case r = 4, $\lambda  $ = 4 and p = 0.3  
with $n_{1} $  = 4.14 , $n_{2} $  = 5.36 and $n_{3} $  = 6.13 ). This  
figure suggests that it might be difficult to obtain experimentally  
a very good relative precision on the power law exponent.}
\label{Fig1}
\end{figure}
\begin{figure}
\caption{Variation of the $\lambda _{L }^{-2} $  (T) 
/ $\lambda _{L }^{-2} $  (0) for an Einstein spectrum for a 
coupling strength $\lambda  $  
= 1 ( short dashed line ), $\lambda  $ = 4 ( dashed line ) 
and $\lambda  $ = 10   
( long dashed line ). The full curve is the BCS result}
\label{Fig2}
\end{figure}
\begin{figure}
\caption{Exponents $n_{1} $  , $n_{2} $ and $n_{3} $  for  r = 4  and  
$\lambda  $ = 2 ( dashed line ), $\lambda  $ = 3 ( open circles ), 
$\lambda  $ = 4 ( full  
lines ), $\lambda  $ = 6 ( filled squares ) and $\lambda  $ = 8 ( dashed lines  
). The lines are just guides for the eye}
\label{Fig3}
\end{figure}
\begin{figure}
\caption{Exponents $n_{1} $  , $n_{2} $ and $n_{3} $  for  r = 6  and  
$\lambda  $ = 2 ( dashed line ), $\lambda  $ = 3 ( open circles ), 
$\lambda  $ = 4 ( full  
lines ), $\lambda  $ = 6 ( filled squares ) and $\lambda  $ = 8 ( dashed lines  
). The lines are just guides for the eye}
\label{Fig4}
\end{figure}
\begin{figure}
\caption{1 $- $ $\lambda _{L }^{-2} $  (T) / $\lambda _{L }^{-2} $  (0) for 
r = 6 , $\lambda  $ = 4 and p = .75 together with its power law approximation  
0.59 $(T/T_{c } $  $)^{2.6} $ over the full temperature range and  
(inset) for $T/T_{c } $ $\leq  $ 0.5 }
\label{Fig5}
\end{figure}
\begin{figure}
\caption{Exponents $n_{1} $  , $n_{2} $ and $n_{3} $  for  r = 8  and  
$\lambda  $ = 2 ( dashed line ), $\lambda  $ = 3 ( open circles ), 
$\lambda  $ = 4 ( full  
lines ), $\lambda  $ = 6 ( filled squares ) and $\lambda  $ = 8 ( dashed lines  
). The lines are just guides for the eye}
\label{Fig6}
\end{figure}
\begin{figure}
\caption{Exponents $n_{1} $  , $n_{2} $ and $n_{3} $  for  r = 16   
and $\lambda  $ = 2 ( dashed line ), $\lambda  $ = 3 ( open circles ), 
$\lambda  $ = 4  
( full lines ), $\lambda  $ = 6 ( filled squares ) and $\lambda  $ = 8 
( dashed   lines ). The lines are just guides for the eye}
\label{Fig7}
\end{figure}
\begin{figure}
\caption{The prefactor of the approximate power law 
for 1 $- $ $\lambda _{L }^{- 2} $  (T) / $\lambda _{L }^{-2} $  (0) : 
r = 4 (left), r = 8 (middle)   
and r = 16 (right); $\lambda  $ = 2 (open diamonds), $\lambda  $ = 3 
(open circles),   
$\lambda  $ = 4 (open squares), $\lambda  $ = 6 (filled squares) and 
$\lambda  $ = 8 (filled  diamonds)}
\label{Fig8}
\end{figure}
\begin{figure}
\caption{$\lambda _{L }^{-2} $  (T) / $\lambda _{L }^{-2} $  (0) 
in the large  
spectral width limit for $\lambda _{2} $  = 3. The coupling strength  
$\lambda _{1} $   takes the values m $\lambda _{2} $ /(10 -m) and the integer  
m goes from 0 to 9. The heavy line corresponds to $\lambda _{1} $  =  
1.28 }
\label{Fig9}
\end{figure}


\begin{references}
\bibitem{remarque}In this respect our procedure is clearly not a good  
one for the results of \cite{hardy} and \cite{lanb}since a clear linear  
dependence is obtained down to low temperature. Anyway these  
experiments do not seem to be compatible with a simple strong  
coupling interpretation with reasonable parameters. The possibility  
of a linear T dependence will be discussed later on in the paper.
\bibitem{disc}We find 0.78 for this T = 0 limit, which is slightly different  
from the corresponding value 0.9 in \cite{carb}.
\end{references}
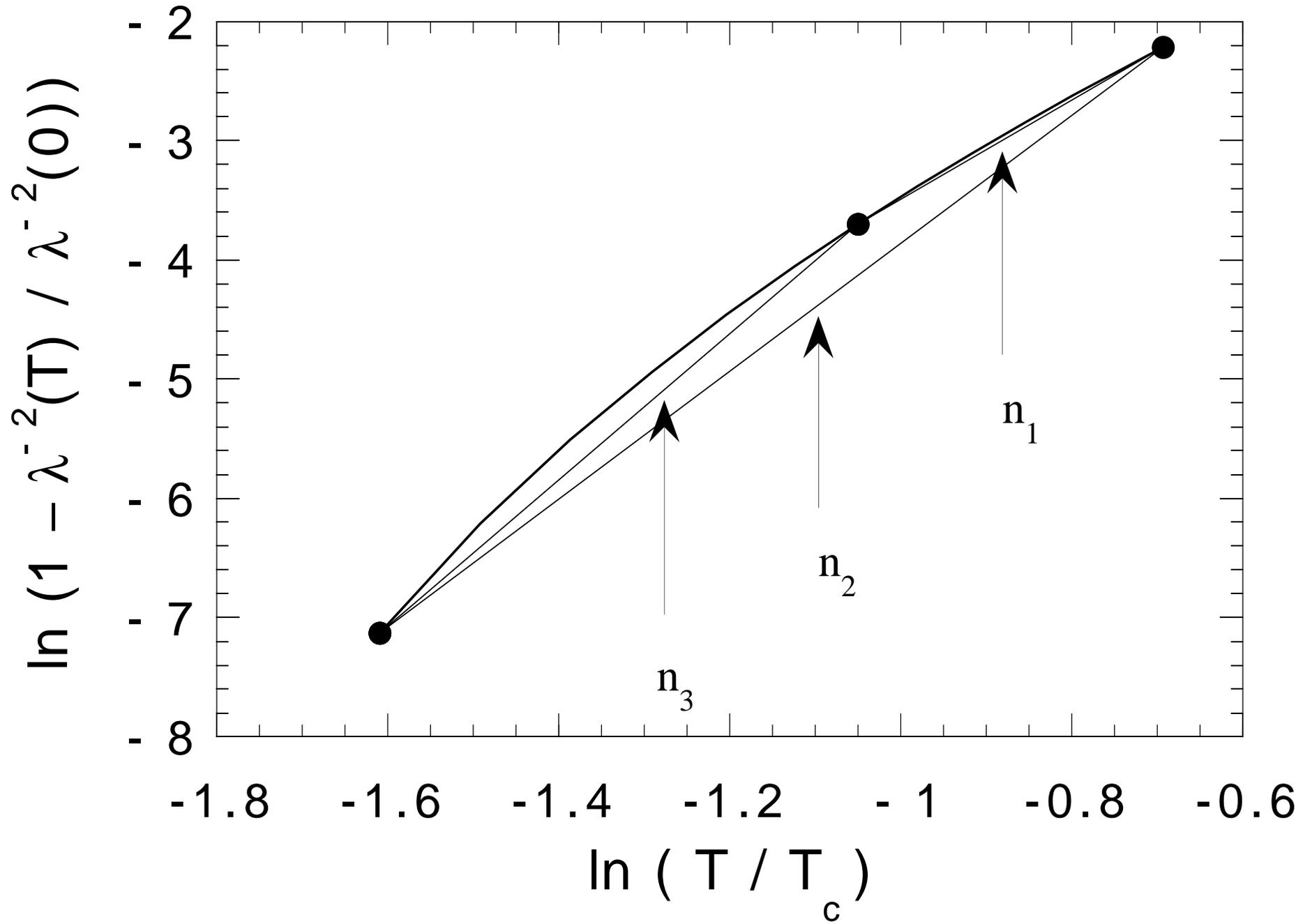
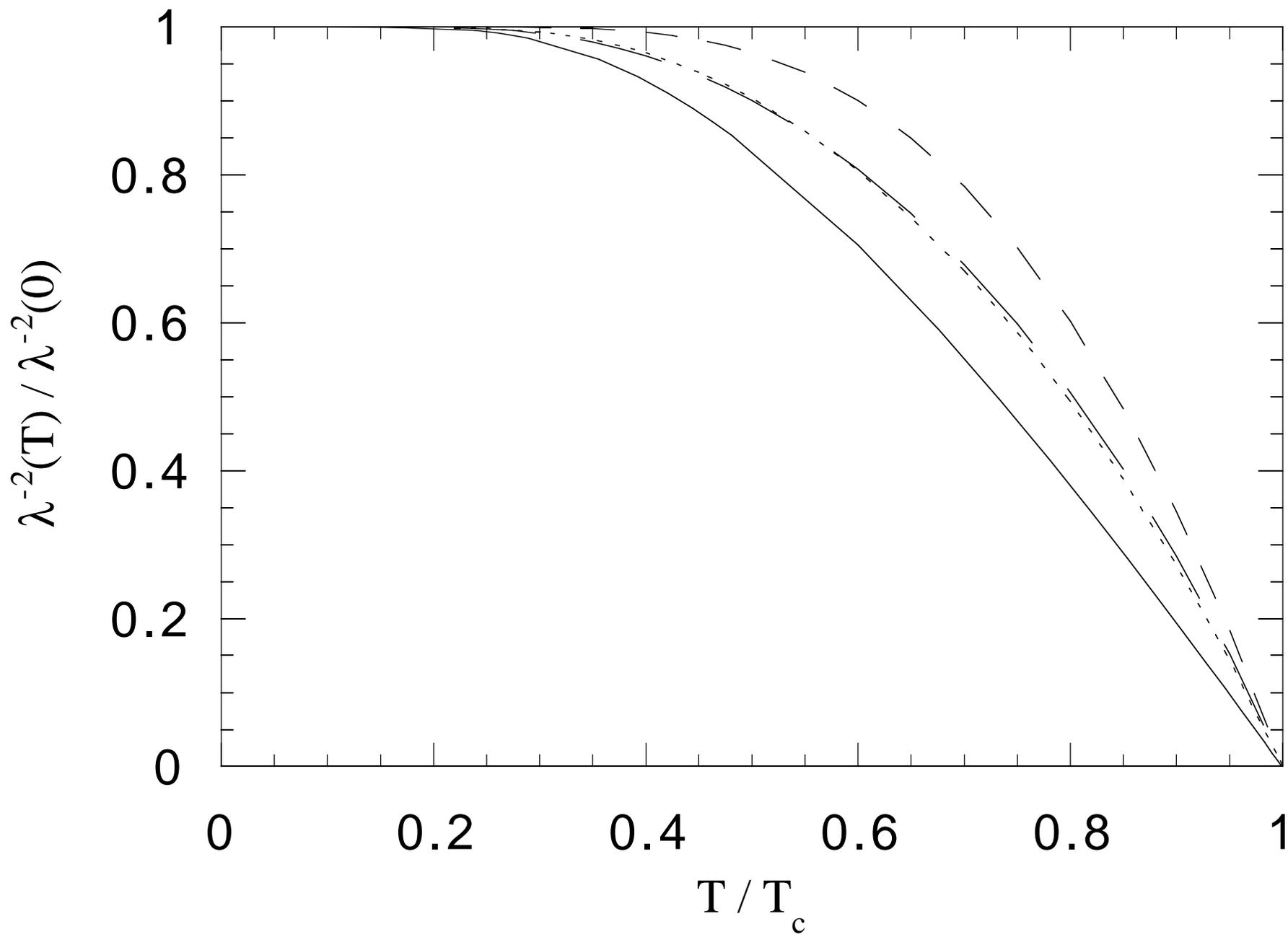
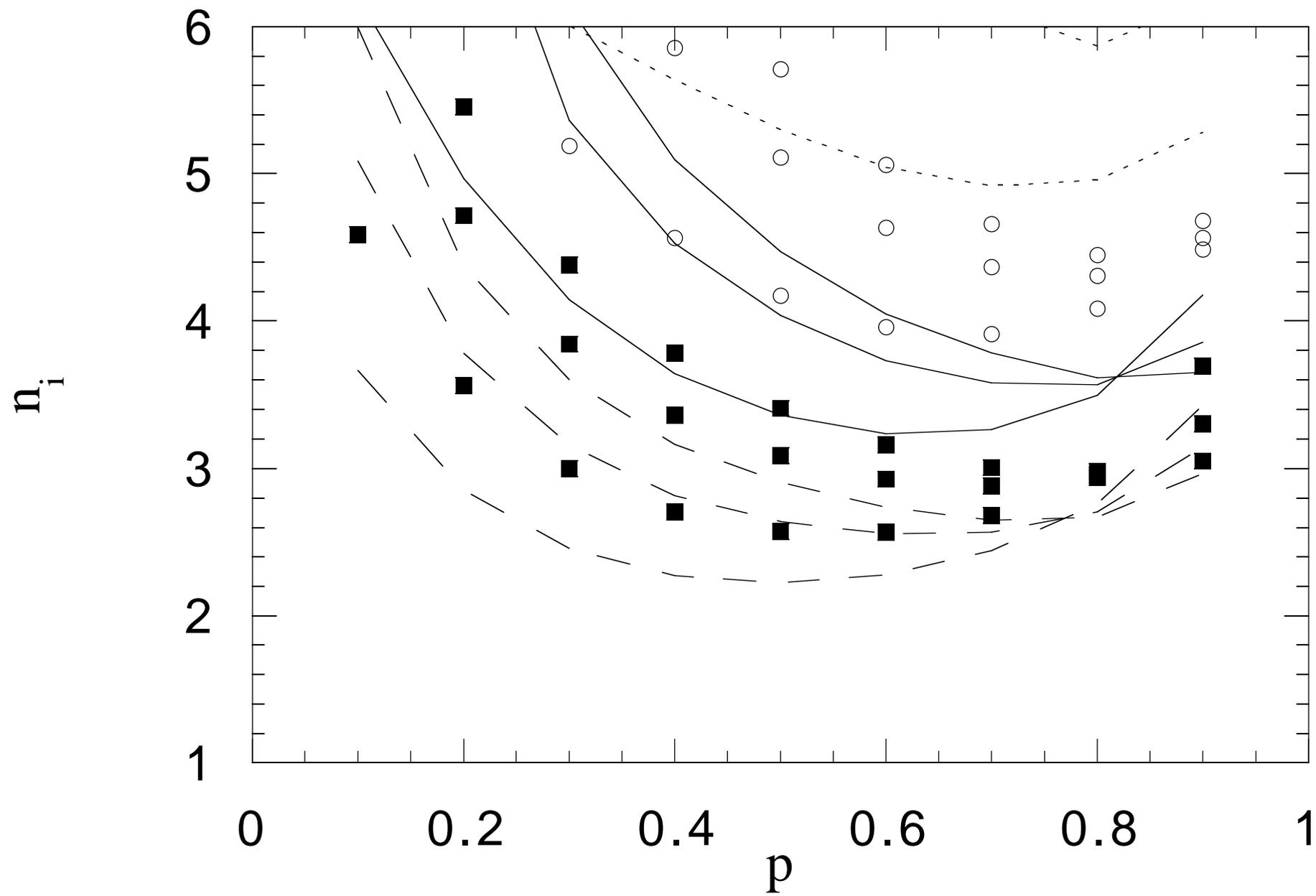
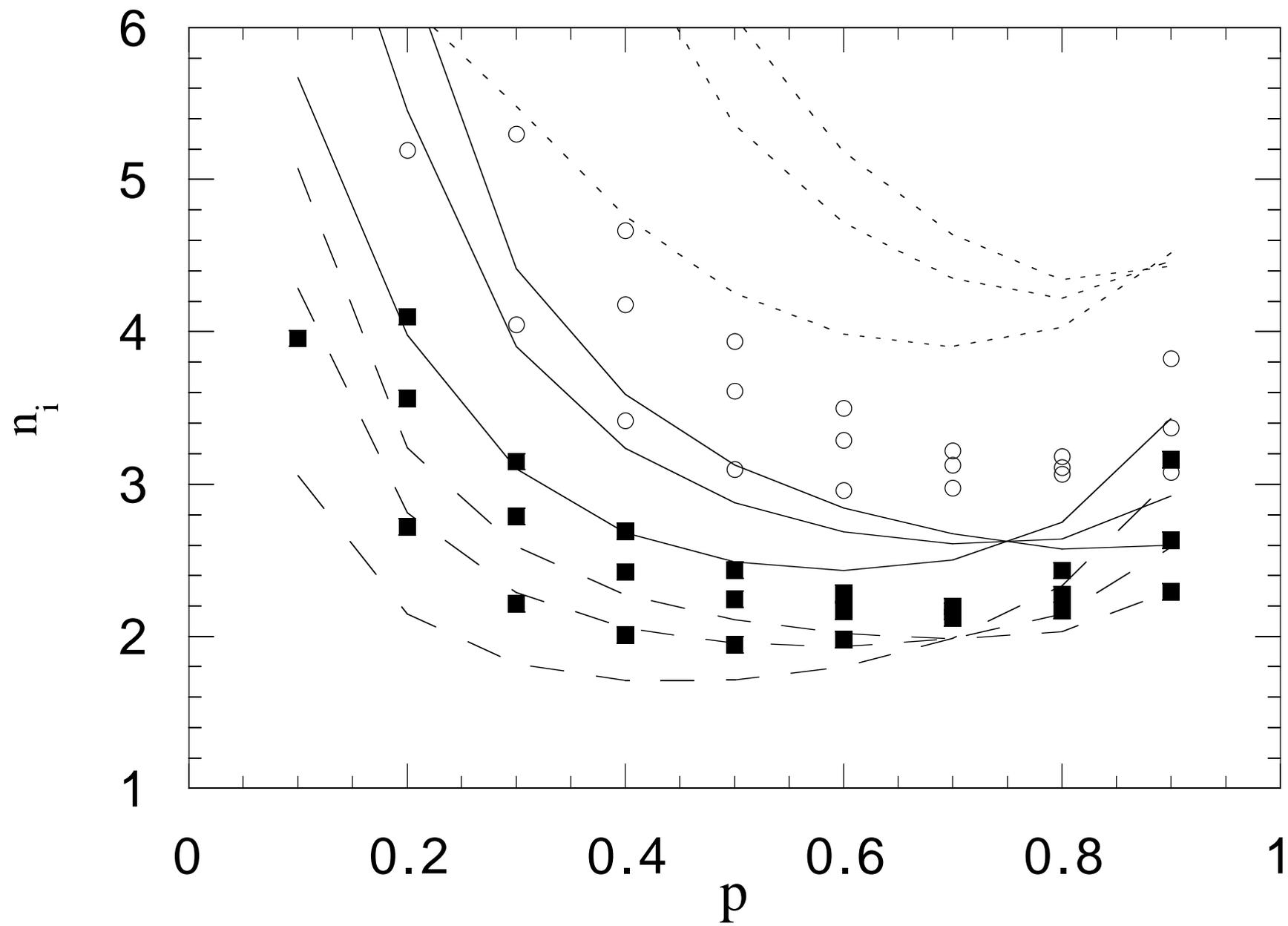
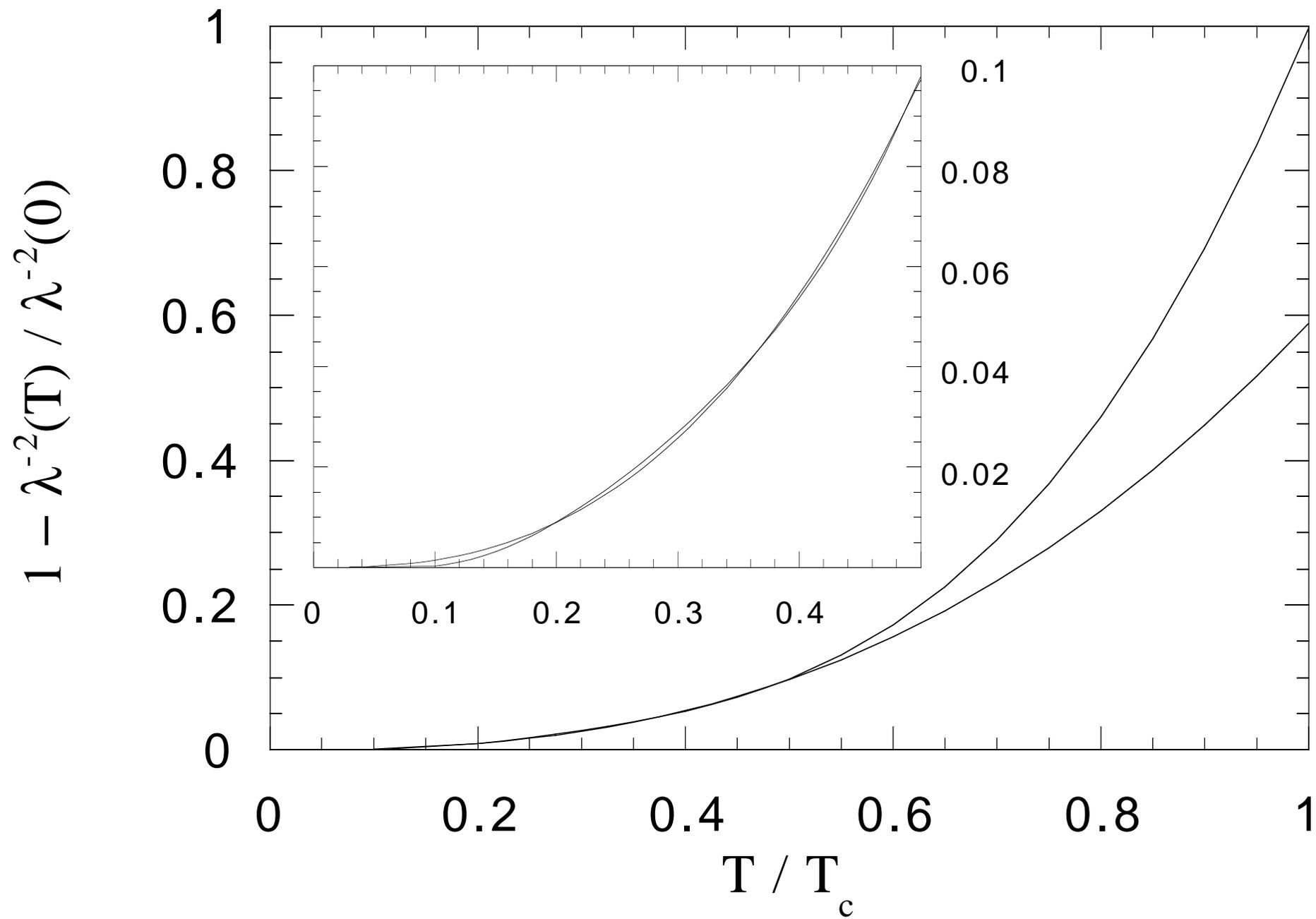
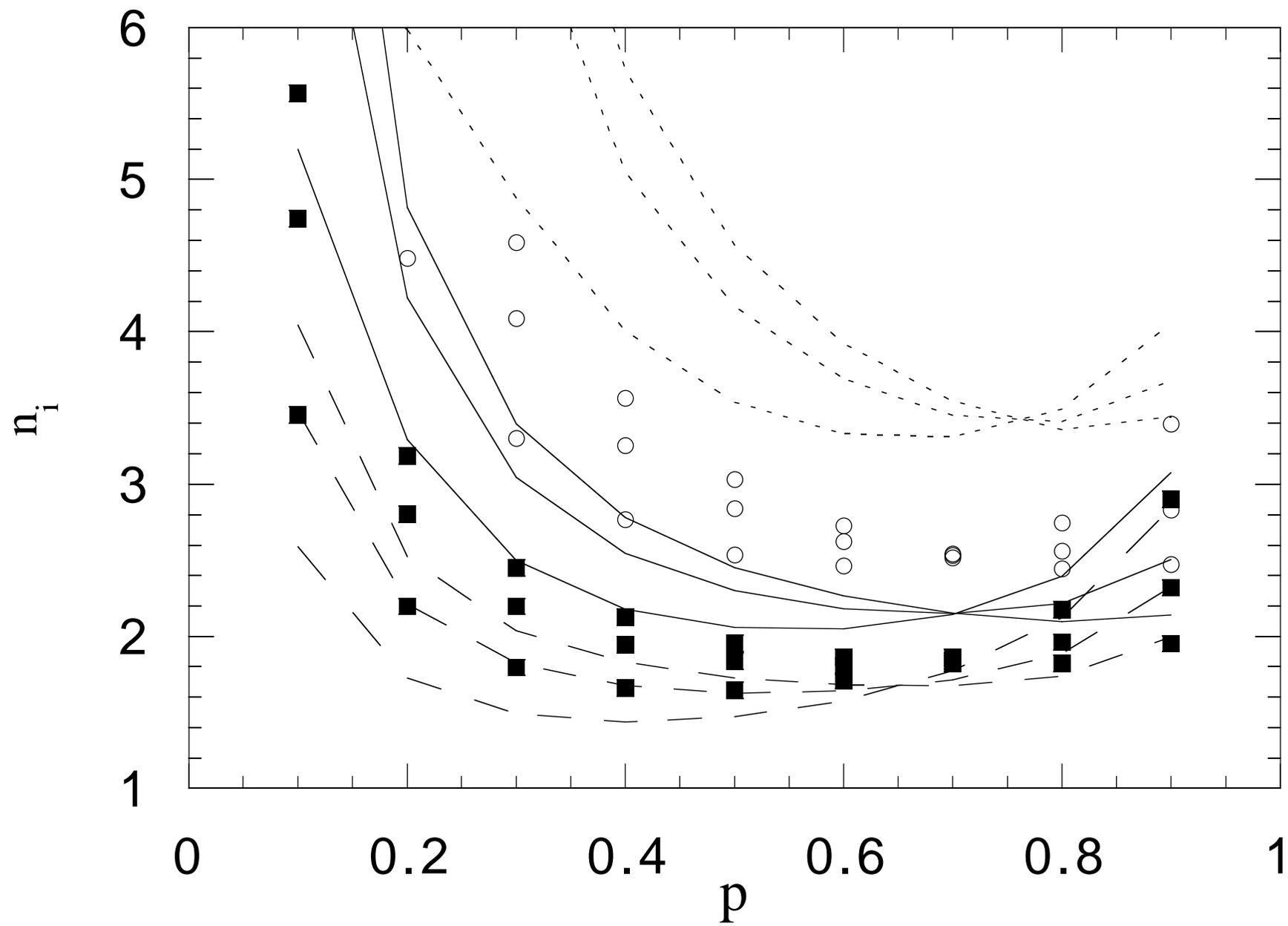
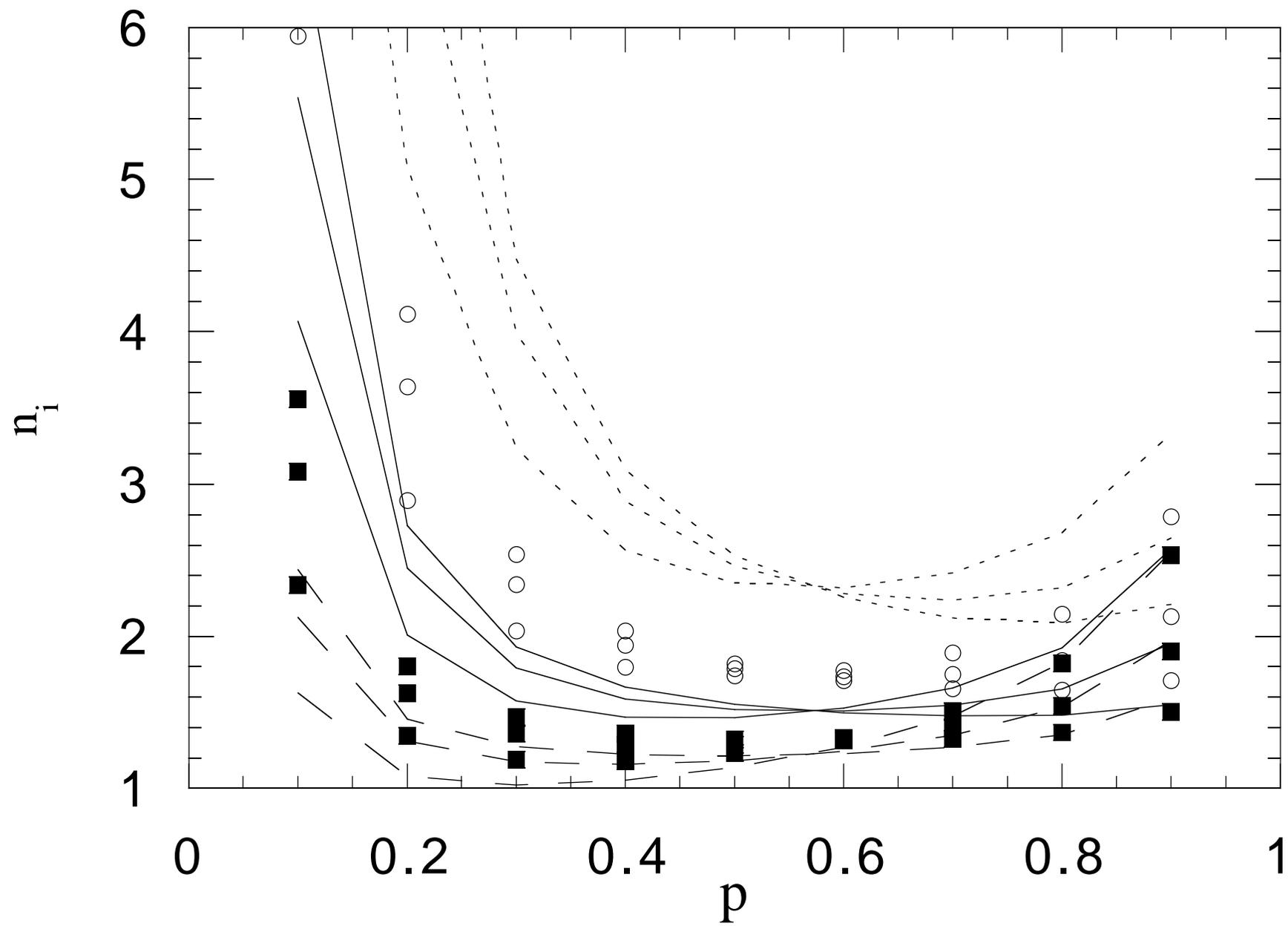
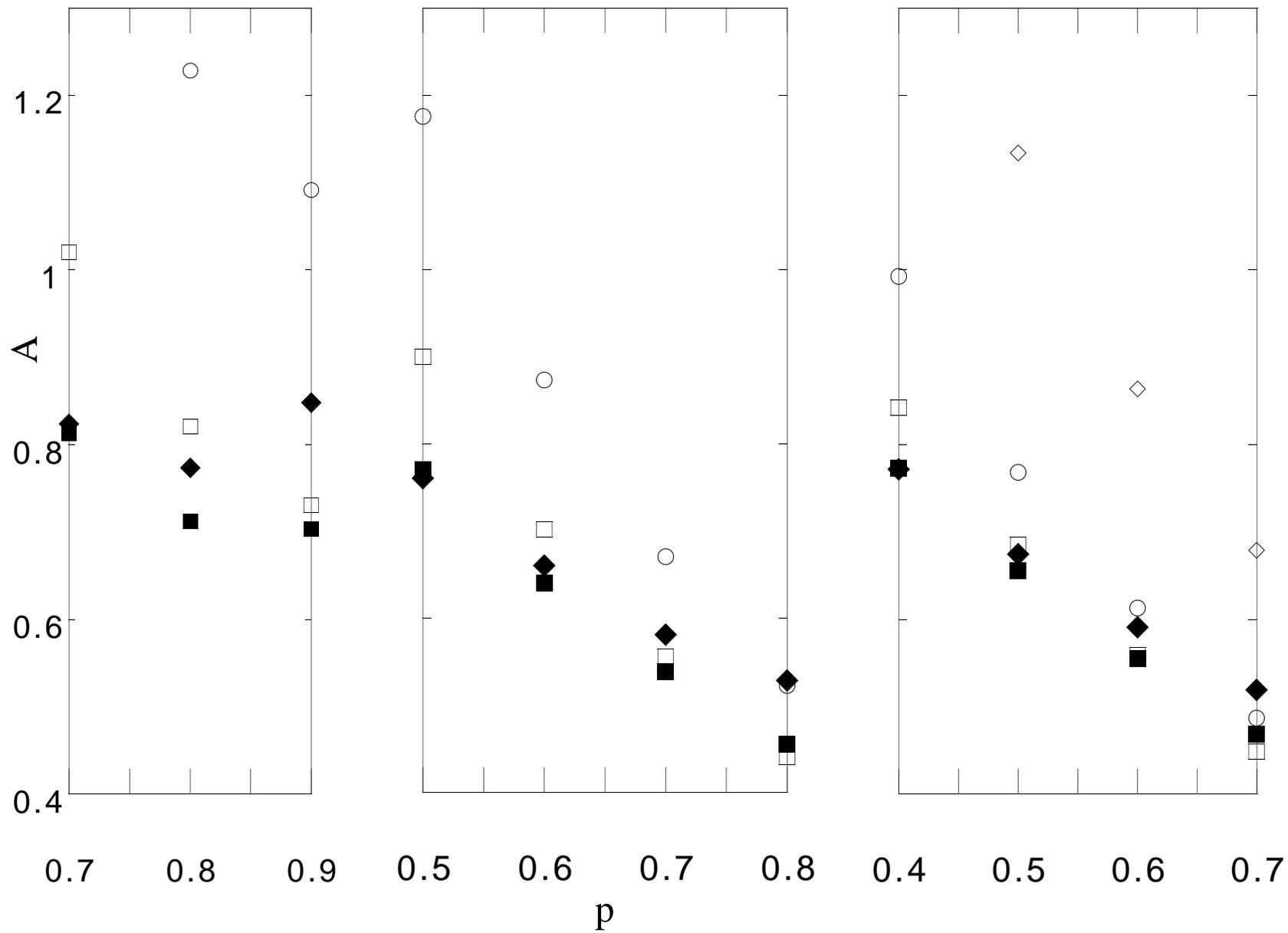
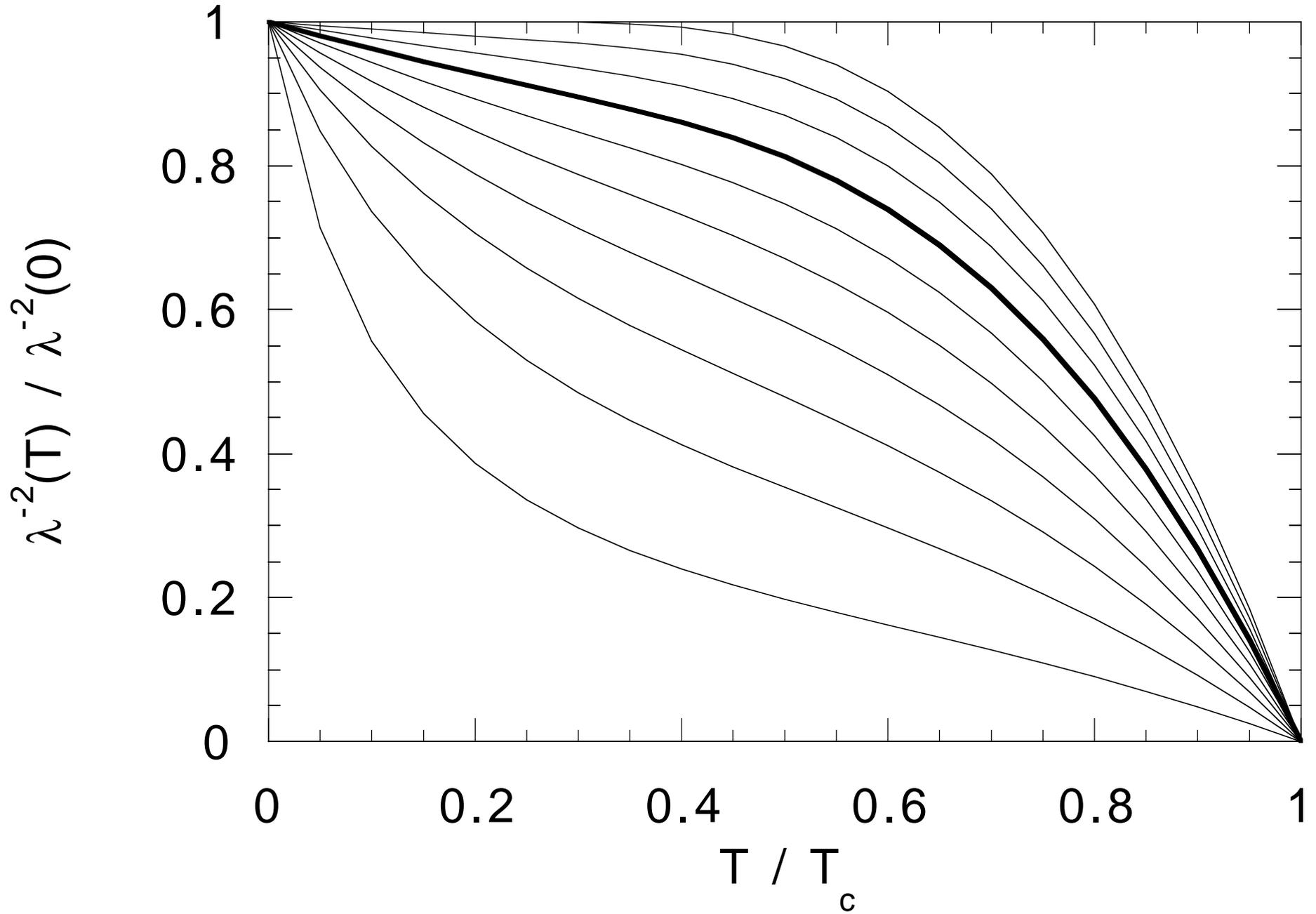
\end{document}